\documentclass[aps,prb,twocolumn,showpacs]{revtex4}
\usepackage{graphicx}
\usepackage{amsmath}
\usepackage{amssymb}

\begin{document}

\title{Spin Dynamics of a Magnetic Antivortex}

\author{Hao Wang}
\author{C.\ E.\ Campbell}
\email{campbell@umn.edu} \affiliation{School of Physics and
Astronomy, University of Minnesota, 116 Church St.\ SE,
Minneapolis, Minnesota 55455, USA}

\date{\today}

\begin{abstract}
We report on a study of the dynamics of a magnetic antivortex in a
submicron, asteroid shaped, permalloy ferromagnet using
micromagnetic simulations. As with vortex states in disk and
square geometries, a gyrotropic mode was found in which a shifted
antivortex core orbits about the center of the asteroid. Pulsed
magnetic fields were used to generate azimuthal or radial spin
wave modes, depending on the field orientation.  The degeneracy of
low frequency azimuthal mode frequencies is lifted by gyrotropic
motion of the antivortex core, and restored by inserting a hole in
the center of the particle to suppress this motion. We briefly
compare the dynamics of the vortex state of the asteroid to the
antivortex. The size dependence of the antivortex modes is
reported.
\end{abstract}


\maketitle

    Recently, there have been substantial efforts to
understand the excitation spectrum of magnetic vortex
structures.\cite{key-1,key-2,key-3,key-4,key-5,key-6,key-7,key-8,key-9,key-10,key-11,key-22,
key-23} There exist two basic types of magnetic vortex structures
in a quasi-two dimensional ferromagnet--a ``circular vortex,"
which we simply call a vortex in the remainder of this report, and
an ``antivortex"--both of which contain a nanometer scale core
area where the magnetization is perpendicular to the plane of the
sample.\cite{key-12,key-13} Cartoons of an antivortex and a vortex
in an asteroid shaped particle are shown in Fig.\ 1. The winding
number of an antivortex is $-1$ and $+1$ for a vortex. In addition
to a fundamental interest in understanding the physics of these
simple structures, the singular spin configurations in magnetic
vortices and their dynamics suggest possible applications in spin
logic operations.\cite{key-4,key-14,key-15} Much research has been
done on a vortex in a circular disk, both theoretically and
experimentally.\cite{key-2,key-3,key-5,key-6,key-7,key-9,key-10,key-11}
Two classes of excitations in the circular disk have been
identified. One is associated with the gyrotropic motion of the
core about its equilibrium position with a frequency lower than 1
GHz for micron sized disks.\cite{key-2,key-3} The other type
consists of spin wave modes at higher
frequenies.\cite{key-5,key-6,key-7,key-9,key-10,key-11} The
excitations of a vortex state in particles with a ``cross"
structure and in rectangular particles have also been observed;
these differ from circular particles both because of the lower
symmetry of the particle and because of the Landau domain
structure around the vortex.
\cite{key-1,key-3,key-4,key-8,key-22,key-23}

We are not aware of research on the dynamics of an antivortex, but
such a study may be useful in understanding the dynamical
properties of a complex multi-vortex structure such as in cross
tie walls and in long
particles.\cite{key-13,key-16,key-17,key-18,key-19}  Since an
antivortex has been found experimentally in an asteroid shaped
permalloy particle (see Fig.\ 1),\cite{key-13} we focus on that
system in this report.

We found that there is also a metastable vortex state at all sizes
studied, shown in Fig.\ 1 (b), as well as two metastable or stable
single domain states; the latter will not be discussed in this
note. Both the vortex and antivortex states are sufficiently
stable to sustain the weak perturbations that we applied without
evolving into another metastable state.

Considering the shape complexity of an asteroid, here we report
only on micromagnetic simulations of the antivortex dynamics in
this particle.\cite{key-20} A gyrotropic mode was observed and two
kinds of spin wave modes were excited by applying pulsed fields at
different angles to the sample plane. The size dependence of the
dynamic excitations was also systematically studied.
%
\begin{figure}
\centerline{\includegraphics{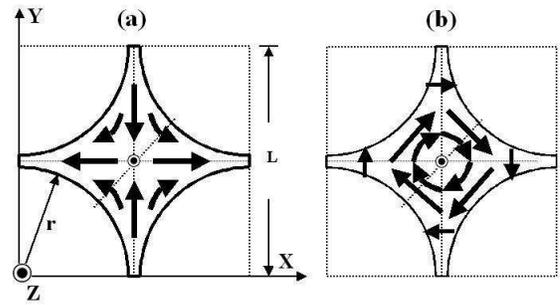}} \caption{Asteroid
particles with a magnetic antivortex (a) and vortex (b). The solid
arrows represent the direction of equilibrium magnetization inside
the asteroid sample.  The core region is indicated by the small
circle about the center, and the dot at the center signifies a
core pointing upward out of the plane.}
\end{figure}

Asteroids were simulated with lateral size $L$ ranging from 200 nm
to 1 $\mu$m and thickness $t$ ranging from 10 to 30 nm. The
circular edges of the particles had a radius of $r=(96/200)L$.
Typical permalloy material parameters were used, with saturation
magnetization $M_{s}=800$ emu/cc, exchange stiffness constant
$A=1.05$ $\mu$erg/cm, and gyromagnetic ratio $\gamma=17.6$ MHz/Oe.
For most simulations, a damping parameter of $\alpha=0.04$ was
used. The cell size for the discretized simulations was chosen to
be 4 nm or 5 nm, which is shorter than the exchange length of 6 nm
with the given material parameters.
%
\begin{figure}
\centerline{\includegraphics{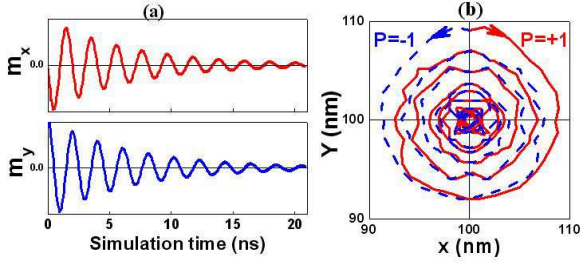}} \caption{(a) Time
evolution of the x and y  magnetization components  in gyrotropic
motion of an initially displaced core moving in remanence. (b)
Trajectories of the antivortex core around its equilibrium
position. The solid, clockwise helix is for core polarity $P=+1$,
while the dashed, counter-clockwise path is for core  $P=-1$.
$L$=200 nm and $t$=20 nm.}
\end{figure}

Because of the out-of-plane magnetization of the antivortex core,
one may expect that a gyrotropic mode will appear once the core is
shifted away from its equilibrium position and
released.\cite{key-2} The core was shifted by applying a constant
in-plane magnetic field. After the core stabilized at an
off-center position, the applied field was removed. The remaining
gyrotropic force on the antivortex core caused it to orbit about
its equilibrium position, spiraling back to the center due to the
damping. Fig.\ 2 shows the time evolution of the two in-plane
components of the average magnetization and the trace of the
antivortex core for a 200x200x20 nm asteroid particle. The damped
periodic behavior of the magnetization shows a gyrotropic mode
with frequency 0.5 GHz. For an antivortex structure, the
gyrotropic vector is proportional to the polarity of its
core.\cite{key-2} Therefore, changing the polarity of an
antivortex core causes the gyrotropic orbit of the core to be in
the opposite direction, as shown in Fig.\ 2(b), where $P=+1$
corresponds to the magnetization of core pointing up and $P=-1$
signifies pointing down.

Spin wave modes with higher frequencies may be excited by applying
magnetic tipping pulse fields. We used Gaussian shaped pulses with
a width of 30 ps and an amplitude of 5 Oe. To couple to a variety
of spin waves, pulsed fields both in the plane of the particle and
perpendicular to its plane were used. As one would expect from the
additional torque exerted on the magnetic moments, in-plane pulses
primarily activate spin wave modes with in-plane wave vectors
curling around the center, i.e. azimuthal-like modes, while a
perpendicular pulse couples most strongly to spin wave modes with
the in-plane wave vectors along the radial direction, i.e.
radial-like modes. Following the Fourier transform techniques of
Ref. \onlinecite{key-9}, we obtained a two-dimensional complex
spectrum, with information about both the spectral amplitude and
the spectral phase. The resonances in the power spectrum indicate
eigenmode frequencies.

The power spectrum resulting from an in-plane pulsed field applied
to a 500x500x20 nm sample along $\widehat{x}$ direction is shown
in Fig.\ 3 (a) for several cases. Of course, the spectra for the
two oppositely polarized cores are identical.  The low frequency
resonance peak in the power spectrum of each of these corresponds
to the core gyrotropic mode, while the resonances in the higher
frequency region are spin wave modes that couple to the uniform
pulse. The spectral details for each of the spin wave modes are
shown in the Fig.\ 3(b) for the $P=+1$ case. The two lowest spin
wave modes with frequencies of 3.8 GHz and 4.7 GHz are clearly
azimuthal modes. They are mainly localized in an ``X" shape region
along the diagonal directions of the asteroid, where a spin wave
well structure is formed due to the relative smallness of the
internal field. The spectral phase of both modes continuously
wraps through $2\pi$ around the azimuthal direction in the
vicinity of the center of the asteroid but with opposite
chiralities, indicating that these two modes are traveling waves
with an angular mode number $|m|=1$. The other excited spin wave
modes with higher frequencies have weight distributed through the
entire asteroid and their spectral phases show that they have a
mixture of both azimuthal and radial characteristics.
%
\begin{figure}
\centerline{\includegraphics{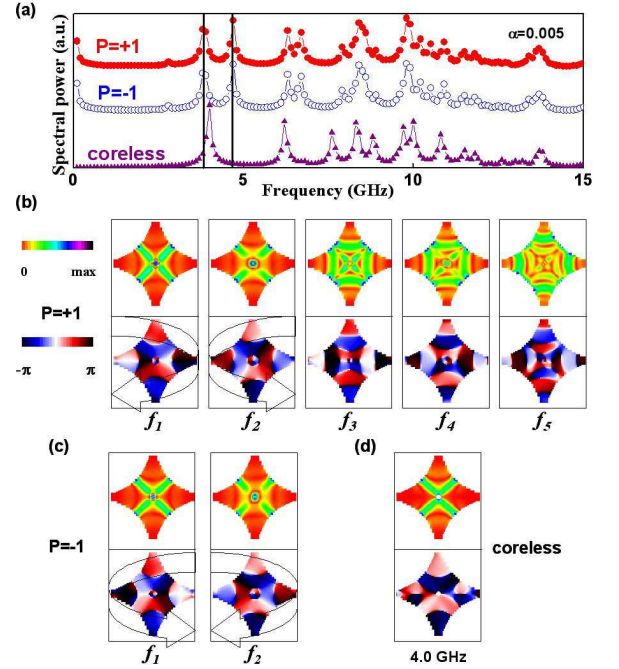}} \caption{(Color online)
(a) Power spectra for 500x500x20 nm asteroids in the antivortex
state. The top pair are for asteroids with opposite antivortex
core polarizations, labeled by $P$. The bottom spectrum is for an
asteroid with a 20 nm hole at the center (coreless). (b) Spectral
amplitude (top) and spectral phase (bottom) at resonance for the
first five azimuthal spin wave modes in the $P=+1$ case at the
resonant frequencies $f_{1}$=3.8 GHz, $f_{2}$= 4.7 GHz,
$f_{3}$=6.3 GHz, $f_{4}$=6.7 GHz and $f_{5}$=8.4 GHz of figure
(a). (c) Spectral amplitude and phase images of the lowest two
azimuthal modes at 3.8 GHz (left) and 4.7 GHz (right) for the
$P=-1$ polarity. (d) Spectral amplitude and phase image of the
lowest azimuthal mode at 4.0 GHz for the coreless asteroid. In (b)
and (c), the chiralities of the two lowest modes with opposite
polarizations are shown by the superimposed broad arrows.}
\end{figure}

One would expect the two lowest azimuthal modes to be degenerate
if the antivortex core stays at the center, which is also
suggested by the similarity of their spectral images. However,
this degeneracy is lifted by the coupling between the azimuthal
spin wave modes and the gyrotropic motion of the core, as was
found in the vortex case.\cite{key-10,key-11} To further explore
the relationship between this splitting and the core's gyrotropic
motion, we chose two 500x500x20 nm asteroids with different core
polarities, the results of which were shown in Fig.\ 3. The two
azimuthal modes for both core polarity cases appear at the same
frequencies as in Fig.\ 3(a). However, the spectral phase images
in Fig.\ 3(b)and (c) reveal a change of the ordering of the
frequency of the two azimuthal modes. For $P=+1$ case in Fig.\
3(b), the 3.8 GHz mode has a clockwise phase chirality while it is
counter-clockwise for the $P=-1$ case. For comparison, we also
simulated an asteroid of the same size but with a central hole of
20-nm diameter in order to remove the antivortex
core.\cite{key-10} Only one resonance peak appears in the range of
3-5 GHz, as shown in the coreless spectrum in Fig.\ 3(a), at a
frequency of 4.0 GHz. The spectral phase of this mode has one
nodal line along the pulse field direction as shown in Fig.\ 3(d),
indicating that the mode oscillates as a standing wave resulting
from two degenerate azimuthal traveling wave modes with opposite
phase chiralities.
%
\begin{figure}
\centerline{\includegraphics{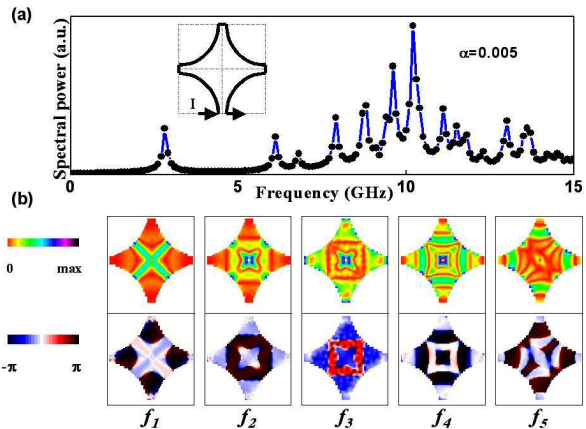}} \caption{(Color online)
(a) Power spectrum for a 500x500x20 nm asteroid after excitation
by an out-of-plane pulse field from a surrounding current. (b)
Spectral amplitude (top) and spectral phase (bottom) at resonance
for first five radial spin wave modes. Their frequencies are
$f_{1}$=2.8 GHz, $f_{2}$=6.1 GHz, $f_{3}$=6.8 GHz, $f_{4}$=7.9
GHz, and $f_{5}$=8.8 GHz.}
\end{figure}

    An out-of-plane pulse field was applied to a 500x500x20 nm sample to
excite radial spin wave modes, results of which are shown in the
Fig.\ 4. Following the experimental setup in Ref.
\onlinecite{key-8}, the pulse simulated was that of an electric
current loop around the asteroid sample as shown in the insert in
Fig.\ 4(a). This pulse will not shift the antivortex core away
from the center. Consequently, the resonances in the power
spectrum in Fig 4(a) are predominantly radial modes and involve no
coupling to the core's gyrotropic motion. The spectral amplitude
and phase of each spin wave mode exhibit a symmetry about the
origin as shown in the Fig.\ 4(b). The lowest radial mode is
mainly distributed in the ``X" shape region as in the azimuthal
mode case. The spectral phase of the lowest mode exhibits nodes
along the four long arms of the asteroid with the variation of the
image color and brightness, while no nodes are along the four
diagonal directions. The amplitudes of higher frequency radial
modes are distributed over the entire asteroid and their phases
show more nodes along both the diagonal and the arm directions of
the asteroid.

    The  difference between the spectrum of the long arm direction
and the diagonal direction of the asteroid can be ascribed to the
different angles between the wave vector and the static
equilibrium internal magnetic field. For radial modes, the wave
vector is perpendicular to the internal field in the diagonal
direction, thus having the characteristics of a Damon-Eshbach
mode(DEM).\cite{key-21} However, along the long arm directions,
the wave vector of radial modes is (anti)parallel to the internal
field, which has magnetostatic backward volume mode (BVM)
character.\cite{key-21} Thus, as a result of the unique spin
configuration of a magnetic antivortex, both the azimuthal and the
radial spin wave modes in the asteroid sample are hybrids of the
DEM and BVM modes.
%
\begin{figure}
\centerline{\includegraphics{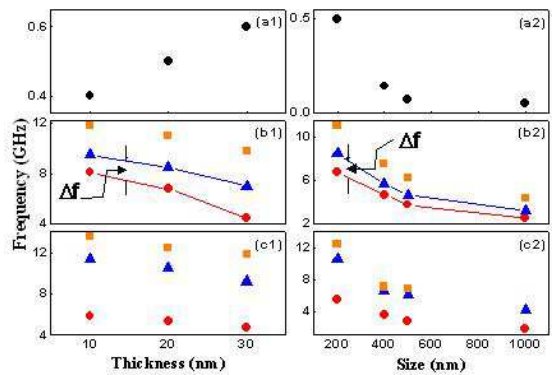}} \caption{Geometry
dependence of dynamic modes of an antivortex in asteroid samples.
The top row of pictures, (a1) and (a2), are results for the core
gyrotropic modes. The middle row of pictures, (b1) and (b2) are
results for the lowest three azimuthal spin wave modes. The bottom
row of pictures, (c1) and (c2) are results for the lowest three
radial spin wave modes. (a1), (b1) and (c1) are for asteroids with
fixed size of 200 nm. (a2), (b2) and (c2) correspond to asteroids
with fixed thickness of 20 nm. For spin wave modes, the different
order of modes are distinguished by different symbols.}
\end{figure}

Fig.\ 5 shows the size dependence of the frequencies of the
various modes obtained in our simulations. The frequency of the
gyrotropic mode increases with the thickness of the asteroid but
decreases with its area. Similar size dependence of the gyrotropic
mode for a vortex in a circular disk has been found
analytically.\cite{key-2} Here, if we treat the asteroid sample
topologically as a deformed circular disk, qualitatively we can
understand the size-dependence of the gyrotropic mode in an
antivortex structure. The frequency of both azimuthal and radial
spin wave modes decreases when either the thickness or the
in-plane size of the asteroid increases. The drop of the frequency
of spin wave modes can be explained by different mechanisms. On
the one hand, an increase of just the in-plane size will
effectively reduce the in-plane wave vector, with a consequent
decrease of the frequency. On the other hand, an increase of just
the thickness results in a lower static internal field while the
in-plane wave vector remains the same. This lower static magnetic
field will shift the dispersion curve to a lower
frequency.\cite{key-21} Thus, for the same in-plane wave vector,
i.e. the same mode, the frequency will decrease. Note also that
the frequency difference between our two lowest azimuthal modes
has the same trend as the gyrotropic mode. This qualitatively
agrees with our previous argument about the coupling between the
core's gyrotropic motion and the degenerate azimuthal modes since
the resulting splitting of azimuthal mode degeneracy should be
proportional to the frequency of the gyrotropic mode.

Finally we compare some of the antivortex results to the vortex
states in the asteroid.  We found that the gyrotropic frequency
for the vortex state is approximately twice that of the
anti-vortex state for the 200 nm asteroid, and four times as large
for the 1 $\mu$m case. An important qualitative difference is that
the chirality of the path of the antivortex core in gyrotropic
motion is opposite to a vortex with the same core polarity. This
will be significant for the low frequency dynamics of
vortex/antivortex arrays.  This feature is simply understood in
terms of the Thiele equation for the gyrotropic motion of the
vortex core, in which the gyrotropic vector is determined by a
product of the core's polarity and its winding number.\cite{key-2}

A comparison of the spin waves in these different systems is
problematic. Here we note several features underlying this.
Firstly, the symmetry group of the antivortex state of an asteroid
is $C_{2v}$, while it is $C_4$ in the vortex asteroid, as it is in
the vortex on a square.\cite{key-22}  Thus the underlying symmetry
classification of the spin wave states is significantly different.
Secondly, the trade-off between exchange energy and dipolar energy
is much more favorable to the formation of Landau domains in both
the asteroid and square particle vortex states than in antivortex
states.\cite{key-22,key-23,key-25} We find that the vortex state
in the larger area asteroids contains well-defined triangular
Landau domains with domain walls along the long axes of the
asteroid, just as in the vortex state in square
particles.\cite{key-8,key-22,key-23} Thus there should be
well-defined intra-domain spin wave modes and localized domain
wall modes in the large asteroid vortex state as found in the
square vortex state.  However, we did not find a tendency toward
domains and domain walls in the antivortex state at any size.
Consequently one expects to find the spin wave modes of the
antivortex in asteroids more difficult to categorize than in these
other cases.

In summary, we have studied dynamic excitations of a magnetic
antivortex structure in asteroid samples and identified several
eigenmodes of the system using micromagnetic simulations. The
modes were generated by magnetic pulses similar to those likely to
be used experimentally.  The generation of lower symmetry modes
would require field pulses of different spatial
symmetry,\cite{key-24,key-25,key-22,key-23} which are not included
here. The size dependence of the modes and the interaction between
two different kind of modes have also been discussed.  A brief
comparison of these results to the dynamics of a vortex in
asteroid and square particles is presented.

The authors gratefully acknowledge discussions with Professors
Paul Crowell and E. D. Dahlberg.  This work was supported in part
by the Office of Naval Research under Grant No. ONR
N/N00014-02-1-0815.

\end{document}